\documentstyle[12pt]{article}
\begin{document}
\thispagestyle{empty}
\setcounter{page}{0}

{}~\vfill
\begin{center}
\Huge BRST - BFV analysis of anomalies\\
in bosonic string theory interacting\\
with background gravitational field
\end{center}
\vfill

\begin{flushleft}
{\large BUCHBINDER I.L., MISTCHUK B.R., \\
\it Department of Theoretical Physics, \\
Tomsk State Pedagogical Institute, Tomsk 634041, Russia.}
\vspace{1cm}

{\large PERSHIN V.D.\\
\it Department of Theoretical Physics, \\
Tomsk State University, Tomsk 634050, Russia.}
\end{flushleft}
\vfill

\centerline{\bf Abstract}
\vspace{0.5cm}

The general BRST-BFV analysis of anomaly in the string theory
coupled to background fields is carried out. An exact equation
for c-valued symbol of anomaly operator is found and structure
of its solutions is studied.
\vfill~

\newpage

     1.  All conventional string models are anomalous at the quantum
level and only under certain conditions on parameters of a theory
anomalies can be cancelled.  In theories of a string interacting
with background fields the role of parameters is playing by these
fields and conditions of anomaly cancellation are interpreted as
effective equations of
motion for them [1-4].  Standard covariant method for deriving
equations of motion for background fields in string theories
is based on the principle of quantum Weyl
invariance which demands that renormalized operator of the
energy-momentum tensor trace vanishes.  General structure of
such an operator in the theory of a string coupled to massless
background fields
was studied in details in refs. [5,6] (see also the reviews [7,8]).

     More consistent (from the theoretical point of view) method
of investigation of anomalous theories is formulated
within the framework of BRST - BFV quantization  [9-12]
(see also the book [17]). The BRST - BFV quantization method, or
the method of generalized canonical quantization provides a powerful
and universal approach to formulation of gauge theories.
Condition of anomaly cancellation in this approach consists in
preserving of gauge algebra of the theory at the quantum level.
In the case of string theories it reduces to the nilpotency
condition for the fermionic operator $\Omega$ generating quantum
gauge algebra.

     The explicit form of quantum operators $\Omega$ and $\Omega^2$
in free string theory was constructed and critical values of parameters
(such as space-time dimensions) were obtained from the nilpotency
condition in refs.[13,14](see also [15,16]). Generalization of this
approach to string models with background fields faces problems
connected with very complicated structure of these theories.
A string interacting with background fields is
discribed by a $\sigma$-model type action that is essentially
non-linear and so requires a suitable perturbation scheme.
An attempt to construct such a scheme was undertaken in ref. [18]
where the nilpotency
condition was stadied for bosonic string coupled to massless background
fields
within the expansion both in powers of $\hbar$ and in normal
coordinates. The lowest order was considered in details and
correspondence with covariant approach was also discussed.

     Some general features of the anomaly can be found using only
general properties of the operator $\Omega$ without explicit computation
of $\Omega^2$ and so without construction of any perturbation scheme.
In refs. [19,20] general structure of quantum anomaly in theories
of free bosonic and fermionic strings was derived from the Jacobi
(super)identity for the operator $\Omega$ and dimensional considerations.
It his paper we consider on the same base the theory of bosonic
string coupled to background gravitational field. Being a theory
with non-polynomial interaction, this model provides a number of
new aspects and details, but as we show the structure of anomaly
can be determined up to a function of string coordinates and
two constants.

     2. The closed bosonic string interacting with background
gravitational field is described by the action
\begin{eqnarray}
S = - {1 \over 2} \int \! d^2\sigma \; \sqrt{-g} \;
g^{ab} \partial_a x^\mu \partial_b x^\nu G_{\mu\nu}(x).
\end{eqnarray}
%
Here $G_{\mu\nu}(x)$ is the metric of $d-$dimensional space-time
with coordinates $x^\mu$; $\mu, \nu = 0, 1,...,d-1$. $g_{ab}$ is the
metric of the two-dimensional world sheet of the string, $\sigma^a
=(\tau,\sigma)$
are coordinates on the world sheet; $a, b = 0, 1$.

     The theory possesses two first class constraints [18]
\begin{eqnarray}
L(\sigma)&=&{1 \over 4} G_{\mu\nu} (p_\mu - G_{\mu\alpha}x'^\alpha)
(p_\nu - G_{\nu\beta}x'^\beta ),
\nonumber
\\
\bar L(\sigma)&=&{1 \over 4} G_{\mu\nu} (p_\mu + G_{\mu\alpha}x'^\alpha)
(p_\nu + G_{\nu\beta}x'^\beta ),
\end{eqnarray}
satisfying Virasoro algebra (in terms of Poisson brackets)
\begin{eqnarray}
& &\left\{ L(\sigma) \, L(\sigma') \right\} = -
\left ( L(\sigma)+L(\sigma')\right )\delta^{\prime}(\sigma-\sigma'),
\nonumber
\\
& &\left\{\bar L(\sigma) \,\bar L(\sigma') \right\} =
\left (\bar L(\sigma)+\bar L(\sigma')\right )\delta^{\prime}(\sigma-\sigma'),
\nonumber
\\
& &\qquad \qquad \left\{ L(\sigma) \,\bar L(\sigma') \right\} = 0 .
\end{eqnarray}
%
$p_\mu(\sigma)$ are  momenta
conjugated to $x^\mu(\sigma)$ and $x'^\mu = \partial_\sigma x^\mu$.

      Let us introduce the canonical BRST charge
\begin{equation}
\Omega = \int\limits_0^{2\pi} d\sigma \; \biggl( L(\sigma)\eta(\sigma) +
\bar L(\sigma') \bar\eta(\sigma') -
{\cal P}(\sigma) \eta(\sigma) \eta'(\sigma)
+ \bar{\cal P}(\sigma') \bar\eta(\sigma') \bar\eta'(\sigma') \biggr).
\end{equation}
%
Here $\eta, {\cal P}$ and $\bar\eta, \bar{\cal P}$ are ghost fields
corresponding to constraints $L$ and $\bar L$ respectively.
The charge $\Omega$ is a generator of BRST transformation
$\delta B = \{ B,\;\Omega\}\epsilon $ where $\epsilon$ is a
fermionic parameter and $B$ is an arbitrary functional on extended
phase space. In the case of the theory under consideration it leads
to the following transformation properties for the fundamental
variables:
\begin{eqnarray}
\delta\eta&=& - \eta \; \eta' \;\epsilon ,
\nonumber
\\
\delta\bar\eta&=& \bar\eta \; \bar\eta' \;\epsilon,
\nonumber
\\
\delta x^\mu&=&{1 \over 2} \; (\eta Y^{\mu} + \bar\eta \bar Y^{\mu})\;
\epsilon,
\nonumber
\\
\delta p_\mu&=&
{1 \over 2} \; \partial_\sigma (\bar\eta \; \bar Y_{\mu} -
\eta \;Y_{\mu}) - {}
\nonumber
\\ & &
{} - {1 \over 4} \; \partial_\mu G_{\alpha\beta}Y^{\alpha} Y^{\beta}
(\eta + \bar\eta),
\nonumber
\\
\delta Y^\mu&=&- (\eta Y^{\mu})' -
{1 \over 2} \; \Gamma^{\mu}_{\alpha\beta} Y^{\alpha} \bar Y^{\beta}(\eta +
\bar\eta),
\\
\nonumber
\delta \bar Y^\mu&=&(\bar\eta \bar Y^\mu)' -
{1 \over 2} \; \Gamma^{\mu}_{\alpha\beta} Y^{\alpha} \bar Y^{\beta}(\eta +
\bar\eta)
{}.
\end{eqnarray}
%
Here $\Gamma^\mu_{\alpha\beta}$ are  Christoffel symbols corresponding
to the metric $G_{\mu\nu}$ and
\begin{eqnarray}
Y^\mu = G^{\mu\nu}p_\nu - x^{\prime\mu},
\qquad
\bar Y^\mu = G^{\mu\nu}p_\nu + x^{\prime\mu}.
\end{eqnarray}
%
In terms of $Y^\mu,\bar Y^\mu$ the constraints (2) have the form
\begin{eqnarray}
L = {1 \over 4} \; G_{\mu\nu}Y^{\mu}Y^{\nu},\qquad
\bar L = {1 \over 4} \; G_{\mu\nu} \bar Y^{\mu} \bar Y^{\nu},
\nonumber
\end{eqnarray}
where $Y_\mu =  G_{\mu\nu}Y^{\nu},\; \bar Y_\mu =  G_{\mu\nu}\bar Y^{\nu}$.
The fields $Y^\mu$ and $\bar Y^\mu$
are the generalization of known  Fubini fields very convenient for
further analysis and will be.

      3. Canonical quantization of the theory (1) was considered in ref.[18].
Operators of the constraints $L$ and $\bar L$ represent
complicated functions of creation and annihilation operators for
oscillating modes of the string
and coordinate and momentum operators for zero modes. We will assume that the
operators $L$ and $\bar L$ are expressed in some normal form
(for example, Wick
ordering for oscillating modes and Weyl ordering for zero ones [12, 18]).
The operator $\Omega$ is given by the eq.(4) where $L$ and $\bar L$ are
operators of the constraints and $\eta, {\cal P}$,
$\bar\eta, \bar{\cal P}$ are
ghost operators satisfying known commutation relations.

      Within the BRST-BFV method the theory is anomalous
if the following relations are fulfilled
\begin{eqnarray}
[\;\Omega,\; \Omega\;]&=&A \neq 0, \\
\left [ \; \Omega,\; H_T \;\right ]&=&A_H \neq 0.
\end{eqnarray}
Here [ , ] is supercommutator, $H_T=H_0+[\Psi,\Omega]$
is the total hamiltonian and $\Psi$ is a gauge fermion [9-12].
If $A$ and $A_H$ are equal to zero the theory is called anomaly free.
Operators $A$ and $A_H$ are called anomalies.

Obviously,
the operator $\Omega$ must obey the operatorical (super)Jacobi identity
regardless of whether the theory is anomalous or not:
\begin{eqnarray}
[\; \Omega,\; [ \;\Omega ,\; \Omega \;]] = 0
\end{eqnarray}
%
The eqs.(7,9) allow to write the equation for anomaly $A$:
\begin{eqnarray}
[\;\Omega,\; A\;] = 0.
\end{eqnarray}
As for the eq.(8), in string theories $A_H$ does not represent an
independent anomaly because the canonical hamiltonian $H_0$ equals zero
and so
\begin{eqnarray}
H_T=[\Psi,\Omega].
\end{eqnarray}
Therefore, the vanishing of $A_H$ (8) is a consequence of the vanishing
of $A$ and the identity (9).

      We will show that the eq.(10)
is very powerful restriction on the form of anomaly $A$ and allows to  write
down the $A$ explicitly up to an arbitrary function having a definite
tensor structure  and depending only on $x^{\mu}$.
      Using eq.(4,10) one can get the anomaly $A$ as follows
\begin{eqnarray}
& &A = \int d\sigma d\sigma' \;\biggl \{ \eta(\sigma)\eta(\sigma')
([ L(\sigma) , L(\sigma') ] + i\hbar( L(\sigma) + L(\sigma'))\delta'(\sigma-
\sigma')) + {}
\nonumber
\\ & & \qquad
{} + \bar\eta(\sigma) \bar\eta(\sigma')([ \bar L(\sigma), \bar L(\sigma') ] -
 i\hbar( \bar L(\sigma) + \bar L(\sigma'))\delta'(\sigma-\sigma')) + {}
\nonumber
\\ & & \qquad
{} + i \hbar^2 {13 \over {12\pi}} (\eta(\sigma)\eta(\sigma') -
\bar \eta(\sigma)\bar \eta(\sigma'))
\delta^{\prime\prime\prime} (\sigma - \sigma') + {}
\nonumber
\\ & & \qquad
{} + (\eta(\sigma) \bar\eta(\sigma') -  \bar\eta(\sigma) \eta(\sigma'))
 [ L(\sigma), \bar L(\sigma') ] \biggr \}
\end{eqnarray}
%
We see that the form of anomaly is defined by the commutators  $[L, L], [\bar
L,
\bar L]$ and $[L, \bar L] $.

     It is convenient to carry out further analysis in terms of symbols of
operators.
Let $B$ and $C$ are some operators depending on operators of canonical
variables
and written in some normal form. Any operator can be associated with
a $c -$valued function of classical arguments $ \Gamma^M \equiv
(x^{\mu}(\sigma), p_{\mu} (\sigma)) $ called the normal symbol
of the operator (see for details [12,21]).  Let $B(\Gamma)$ and
$C(\Gamma) $ be symbols of the operators $B$ and $C$. Symbol
corresponding to the product $BC$ is called $*$-product of the
symbols $B(\Gamma)$ and $C(\Gamma) $ and has the form
\begin{eqnarray}
B(\Gamma)*C(\Gamma)= \left. {\rm exp} \left\{ \Gamma_1^M \Gamma_2^N
{\delta \over \delta \Gamma_1^M} {\delta \over \delta \Gamma_2^N} \right\}
B(\Gamma_1) C(\Gamma_2) \right|_{\Gamma_1= \Gamma_2= \Gamma}
\end{eqnarray}
%

Here $ \Gamma'^M \equiv (x^{\mu}(\sigma'), p_{\mu}(\sigma')) $;
the sum over $M, N$ implies the summation over $\mu, \nu$ and
integration over $\sigma, \sigma'$.  The contraction in (13)
corresponds to the choosed type of ordering.
 The symbol of commutators $[B,
C]$ is  $B(\Gamma) * C(\Gamma') - C(\Gamma') * B(\Gamma) $ and is
called $*$-commutator.

      Let us consider $*$-commutator corresponding to the commutator
$[L(\sigma), L(\sigma')]$. Taking into account eq.(13) one obtains
\begin{eqnarray}
L(\sigma)*L(\sigma') - L(\sigma')*L(\sigma) =
i \hbar \left \{ L(\sigma),\;L(\sigma')\right \} + A^{(1)}(\sigma,\sigma').
\end{eqnarray}
%
The function $A^{(1)}(\sigma, \sigma')$ appends when we expand the
exponential in eq.(13) in series in power of contractions.
Since every contractions contains $\hbar$, $*$-commutator should be in
general case an infinite series in powers of $\hbar$. The first term in
(13) is proportional to classical Poisson bracket, the following terms
correspond to loop corrections. In free string theory the constraints
are quadratic
function of $\Gamma^M$. It means that the series in $\hbar$ terminetes of the
second order and we get the known result for the free string anomaly. In the
string theory interacting with arbitrary background fields the series for
$A^{(1)}(\sigma, \sigma')$ contains all orders in $\hbar$ and its
straightforward calculation is difficult and complicated enough problem.
Symbols corresponding to commutators  $[\bar L(\sigma),\bar L(\sigma')]$
and  $[L(\sigma),\bar L(\sigma')]$ have structure analogous to (14).

      For further analysis it is useful to take into account the
dimensions of the basic objects of the theory
\begin{eqnarray}
& &\dim x^\mu = 0, \quad \dim p_\mu =1, \quad \dim \eta = \dim \bar
\eta = c,
\nonumber
\\ & &\qquad
\dim {\cal P} = \dim \bar {\cal P} = 1 - c.
\end{eqnarray}
Then
\begin{eqnarray}
\dim L = \dim \bar L = 2, \quad
\dim \Omega = 1 + c, \quad \dim A = 2 + 2c.
\end{eqnarray}

      Let the functions  $A^{(2)}(\sigma, \sigma')$ and  $A^{(3)}(\sigma,
\sigma')$
are given by the equations
\begin{eqnarray}
\bar L(\sigma)*\bar L(\sigma') - \bar L(\sigma')*\bar L(\sigma)&=&
i \hbar \left \{\bar L(\sigma),\;\bar L(\sigma')\right \} +
A^{(2)}(\sigma,\sigma').
\nonumber
\\
L(\sigma)*\bar L(\sigma') - \bar L(\sigma')*L(\sigma)&=&A^{(3)}(\sigma ,
\sigma')
\end{eqnarray}
%
and the functions $A^{(1)}(\sigma, \sigma')$ is given by the eq.(14).
Let ${\cal A}$ is a symbol corresponding to the operator of anomaly $A$.
Taking into account the eq.(12) we obtain
\begin{eqnarray}
{\cal A}&=&\int d\sigma d\sigma'\;\biggl \{
\eta(\sigma)\eta(\sigma')\;[ A^{(1)}(\sigma, \sigma') +
i\hbar^2 {13 \over {12\pi}}\delta''' (\sigma - \sigma')] +
\nonumber
\\& &
{} + \bar \eta(\sigma)\bar \eta(\sigma')\;[
A^{(2)}(\sigma, \sigma') + i \hbar^2 {13 \over {12\pi}}
\delta'''(\sigma - \sigma')] +
\nonumber
\\ & & {} +
[\eta(\sigma) \bar\eta(\sigma') - \bar\eta(\sigma) \eta(\sigma')]\;
A^{(3)}(\sigma,\sigma') \biggr \}.
\end{eqnarray}
%

Here $ \dim A^{1} = \dim A^{2} = \dim A^{3} = 4$ and $ \eta(\sigma),
\bar\eta(\sigma')$ are classical anticommuting functions. The eqs.(14.17)
show that $A^{(1)}(\sigma, \sigma') = - A^{(1)}(\sigma', \sigma),
A^{(2)}(\sigma, \sigma') = - A^{(2)}(\sigma', \sigma)$. It allows to
write
\begin{eqnarray}
A^{(1)}(\sigma,\sigma') &=&
(f_1(\sigma)+f_1(\sigma') \delta'(\sigma - \sigma') +
(f_3(\sigma)+f_3(\sigma') \delta'''(\sigma - \sigma'),
\nonumber
\\
A^{(2)}(\sigma,\sigma') &=&
(g_1(\sigma)+g_1(\sigma') \delta'(\sigma - \sigma') +
(g_3(\sigma)+g_3(\sigma') \delta'''(\sigma - \sigma'),
\end{eqnarray}
%
where arbitrary functions $ f_1,\; g_1,\; f_3,\; g_3 $ have dimensions
\begin{eqnarray}
\dim f_1 = \dim g_1 = 2,
\quad
\dim f_3 = \dim g_3 = 0.
\nonumber
\end{eqnarray}
%
As to the function $A^{(3)}(\sigma, \sigma')$ it has more general
structure
\begin{eqnarray}
A^{(3)}(\sigma,\sigma')&=&{1\over 2}(h_0(\sigma)+h_0(\sigma'))
\delta(\sigma - \sigma') + {1\over 2}
(h_1(\sigma)+h_1(\sigma')) \delta'(\sigma - \sigma') + {}
\nonumber
\\
{}&+&{1\over 2}(h_2(\sigma)+h_2(\sigma')) \delta''(\sigma - \sigma') +
{1\over 2}(h_3(\sigma)+h_3(\sigma')) \delta'''(\sigma - \sigma'),
\end{eqnarray}
%
where arbitrary functions $h_k\;(k = 0,1,2,3)$ have dimensions
$\dim h_k = 3 - k$.

     Using the eqs.(19,20) we can write the anomaly ${\cal A}$ as follows
\begin{eqnarray}
{\cal A}&=&\int d\sigma d\sigma'\;\biggl \{
i\hbar^2 {13 \over {12\pi}}(\eta \eta'''+ \bar \eta \bar \eta''') +
a_0 + a_1 + a_2 + a_3 \biggr \},
\nonumber
\\
a_0&=& 2 \eta \bar \eta\; h_0,
\nonumber
\\
a_1&=&\eta \eta'\;f_1 + \bar\eta \bar\eta'\;g_1 +
2(\eta \bar\eta' - \bar\eta \eta')\;h_1,
\nonumber
\\
a_2&=&2(\eta \bar\eta'' - \bar\eta \eta'')\;h_2,
\nonumber
\\
a_3&=& \eta\;\eta'''f_3 + \bar\eta\;\bar\eta'''g_3 +
2(\eta \bar\eta''' - \bar\eta \eta''')\;h_3.
\end{eqnarray}
%
Thus we obtained the anomaly up to  unknown functions
$h_0$, $f_1$, $h_1$, $g_1$, $h_2$, $f_3$, $h_3$, $g_3$.

     However we have not yet used the eq.(10). Being
rewritten in terms of symbols of operators the eq.(10) leads to
equation of the following type
\begin{eqnarray}
\hat \delta {\cal A} = 0,
\end{eqnarray}
%
where the operator $\hat\delta$ has the structure
\begin{eqnarray}
\hat \delta = \sum\limits_{n=0}^{\infty} \hbar^n \delta_n.
\end{eqnarray}
%
Here $\delta_0 = \delta$ is the operator of classical BRST-transformations
(5). An explicit form of $\delta_n$ at $n>0$ is unessential for our
purposes.

     The functional ${\cal A}$ can be represented as follows
\begin{eqnarray}
{\cal A} = \sum\limits_{n=1}^{\infty} \hbar^{n+1} {\cal A}_n,
\end{eqnarray}
%
where ${\cal A}_n$ are quantum correction to the anomaly. It is easy
to show that the eq.(22-24) lead to
\begin{eqnarray}
\delta {\cal A}_1&=&0
\\
\delta {\cal A}_n&=&G_n, \quad n\geq 2
\\
G_n&=&\sum\limits_{m=1}^n \delta_m {\cal A}_{n-m}.
\nonumber
\end{eqnarray}
%
The eq.(25) is one for the first (one-loop) quantum correction, the eq.(26)
allows to find higher quantum corrections to the anomaly.

      4. Our aim is to investigate the structure of the eqs.(25,26)
solutions. We start with eq.(25) defining the one-loop contribution
to the anomaly.

      Let ${\cal A}_1$ is a solution of the eq.(25). According to the
previous analysis general structure of ${\cal A}_1$ is given by
the eq.(21). It means that to satisfy the eq.(25) we should demand
\begin{eqnarray}
\delta a = \partial_{\sigma} \chi
\end{eqnarray}
%
with some function $\chi, \dim \chi = \dim a-1$. The conditions (27)
impose strict constraints on the functions $h_0, f_1, h_1, g_1, h_2,
f_3, h_3, g_3$.

      Let $\Psi(x,p)$ is any of the above functions with the dimension
$\dim \Psi$. Since the only dimensional quantities are $x'^\mu, p_\mu$ and
$\partial_\sigma$ we can write the $\Psi$ as a polynom of a power $\dim
\Psi$ in $Y_\mu,\bar Y_\mu, Y'_\mu$ and $\bar Y'_\mu$. For example
\begin{eqnarray}
f_1 &=& \beta_{\mu\nu}Y^\mu\bar Y^\nu + \alpha_{\mu\nu}Y^\mu Y^\nu +
 \bar \alpha_{\mu\nu}\bar Y^\mu\bar Y^\nu +
\gamma_\mu Y'^\mu + \bar \gamma_\mu \bar Y'^\mu ,
\end{eqnarray}
%
where $\beta_{\mu\nu}, \alpha_{\mu\nu}, \bar \alpha_{\mu\nu},
\gamma_\mu, \gamma_\mu$ are arbitrary functions of $x$. Analogous
relations can be written for $h_0, f_1, h_1, g_1, h_2, f_3, h_3, g_3$.
In particular, the funcions $f_3, g_3, h_3$ are independent on
$Y, \bar Y$ and their derivatives at all. Taking into account the
eqs.(5) we obtain
\begin{eqnarray}
\delta \Psi = \sum\limits_{n=0}^{\dim \Psi} (\Psi^{(n)} \partial_\sigma^n
\eta + \bar \Psi^{(n)} \partial_\sigma^n \bar \eta )
\end{eqnarray}
%
with some functions $\Psi^{(n)}, \bar \Psi^{(n)}$.

      Let us consider the eqs.(27). It is obviously that the function
$\chi$ should have the structure
\begin{eqnarray}
\chi = \sum (ghost)\:\lambda(x, p),
\end{eqnarray}
%
where $(ghost)$ is a suitable ghost contribution and $\lambda$ are
polynoms of the proper power in $Y, \bar Y, Y', \bar Y'$ with coefficients
depending on $x$. Namely,the power equals $\dim a - \dim (ghost) - 1$. Taking
into account the eq.(27, 30) one gets
\begin{eqnarray}
\sum \delta a = \sum [(ghost)'\;\lambda + (ghost)\;\lambda'].
\end{eqnarray}
%
$\delta a$ can be found on the base of the eqs.(5, 21, 29). It is
obvious that $\delta a$ will have the special structure given by the
right hand side of the eq.(31) only under special restrictions on the
functions $\Psi$. A straightforward but complicated enough
consideration leads to the folloving result
\begin{eqnarray}
 a_0&=&a_2 = 0,
\nonumber
\\
a_1&=&(\eta + \bar\eta)(\eta - \bar\eta)'\beta_{\mu\nu}Y^\mu
\bar Y^\nu,
\nonumber
\\
a_3&=&( \eta\;\eta'''\Lambda + \bar\eta\;\bar\eta'''\bar \Lambda).
\end{eqnarray}
%
Here $\beta_{\mu\nu}$ is an arbitrary function of  $x$  and  $\Lambda,\;
\bar \Lambda$ are the constants. The corresponding  $\chi$ is
\begin{eqnarray}
\chi = i\bar \eta \eta\;(\eta - \bar \eta)'\beta_{\mu\nu}Y^\mu
\bar Y^\nu \epsilon + i(\bar \eta \bar \eta'\bar \eta''\bar
\Lambda - \eta \eta'\eta''\Lambda)\epsilon.
\end{eqnarray}
%

      Thus the one-loop correction to the anomaly has the form
\begin{eqnarray}
{\cal A}_1 = i\hbar^2\;\int d\sigma \;\biggl \{(\eta + \bar\eta)(\eta -
\bar\eta)' \;\beta_{\mu\nu}^{(1)}(x) Y^\mu \bar Y^\nu + \nonumber \\
{}+\eta \eta'''\beta^{(1)} + \bar\eta \bar\eta''' \bar\beta^{(1)} \biggr
\}.
\end{eqnarray}
%
where $\displaystyle \beta^{(1)}={13 \over {12\pi}} +
\Lambda^{(1)}, \bar \beta^{(1)}={13 \over {12\pi}} + \bar \Lambda^{(1)}$
and the index ${}^(1)$ indicates the one-loop
contribution. The eq.(34) is general consequence of the BRST-BFV
quantization procedure.

      5. Let us consider now the eqs.(26) defining the higher quantum
corrections. Writing
\begin{eqnarray}
{\cal A}_n = \int d\sigma a^{(n)}, \quad G_n = \int d\sigma g^{(n)},
\end{eqnarray}
%
we obtain using the eq.(26) $\delta a^{(n)} = g^{(n)} + \partial_\sigma
\lambda^{(n)}$ where $ \lambda^{(n)}$ are some functions. Since
$\delta^2 a{(n)} = 0$ one gets $\delta g^{(n)} = - \partial_\sigma
\delta \lambda^{(n)}$. A solution of this equation can be written as
\begin{eqnarray}
g^{(n)} = g^{(n)}_0 + \partial_\sigma \; \kappa^{(n)},
\end{eqnarray}
%
where $\kappa^{(n)}$ are some functions and
\begin{eqnarray}
\delta g^{(n)}_0 = 0.
\end{eqnarray}
%
The eqs.(35,36) show that
\begin{eqnarray}
G_n = \int d\sigma g^{(n)}_0.
\end{eqnarray}
%

      A solution $g_0^{(n)}$ of the eq.(37) can be found taking into
account that a ghost structure of $g_0^{(n)}$ is defined by the ghost
structure  $\delta a^{(n)}$ and the dimension of $g_0^{(n)}$ eqals
$\dim g_0^{(n)} = 4 + 3c$. The eqs.(18 - 21) allow to write $ a^{(n)} =
a_0 + a_1 + a_2 + a_3$ where $a_k$ are given by (21). Calculating
$\delta a^{(n)}$ we find the general form of $ g_0^{(n)} $
\begin{eqnarray}
g_0^{(n)}&=&\bar\eta\eta\eta'B + \bar\eta\eta\bar\eta'\bar B +
\bar\eta\eta\eta''C +
\bar\eta\eta\bar\eta''\bar C + \bar\eta\eta\eta'''D +
\bar\eta\eta\bar\eta'''\bar D + {}
\nonumber
\\& &
{} + \bar\eta\eta\eta^{(4)} E +
\bar\eta\eta\bar\eta^{(4)} \bar E + \eta\eta'\bar\eta' F +
\bar\eta\bar\eta'\eta'\bar F + \eta\eta'\bar\eta'' H +
\bar\eta\bar\eta'\eta''\bar H + {}
\nonumber
\\& &
{} + \eta\eta'\bar\eta'''S +
\bar\eta\bar\eta'\eta'''\bar S + \eta\bar\eta'\eta'' T
+ \bar\eta\eta'\bar\eta''\bar T + \eta\bar\eta'\bar\eta''R +
\bar\eta\eta'\eta''\bar R + {}
\nonumber
\\& &
{} + \eta\bar\eta'\bar\eta''' U
+ \bar\eta\eta'\eta'''\bar U + \eta\eta'\eta''V
\bar\eta\bar\eta'\bar\eta''\bar V + \eta\eta'\eta'''W
 + \bar\eta\bar\eta'\bar\eta'''\bar W
\end{eqnarray}
%
where $ \Phi \equiv (B, \bar B, C, \bar C, D, \bar D, E, \bar E, F,
\bar F, H, \bar H, S, \bar S, T, \bar T, R, \bar R, U, \bar U, V, \bar
V, W, \bar W $ are some unknown functions  of  $x,  p$  of
definite dimensions.  The  functions  $\Phi$  can  be  presented  as
polynomials in power of monomials $ Y^\mu,  \bar  Y^\mu,  Y'^\mu,
\bar Y'^\mu, Y''^\mu, \bar Y''^\mu, Y'''^\mu,  \bar  Y'''^\mu$.
The number of monomials is defined by dimension  of  $\Phi$  with  unknown
coefficients depending only on $x^\mu$.  The  eq.(39)  and  the
structure of the functions  $\Psi$  allow  to  calculate  $\delta
g_0^{(n)}$ in explicit form. Then  the  eq.(3)  will  lead  to  a
system of the  linear  homogenous  equations  for  the  functions
$\Phi$. The only solution of this system is $\Phi = 0$. It  means that
$g_0^{(n)} = 0$ and, hence, $G_n = 0$ too.

      As a result of above discussion the eqs.(26) defining higher
quantum correction to the anomaly are reduced to
\begin{equation}
\delta {\cal A}_n = 0.
\end{equation}
%
But the equation of this type has been studied in previous section.
Therefore we are able to write down the most general
form of the anomaly
\begin{eqnarray}
{\cal A} = i\hbar^2\;\int d\sigma \;\biggl \{(\eta + \bar\eta)(\eta -
\bar\eta)' \;\beta_{\mu\nu}(x) Y^\mu \bar Y^\nu +  \eta \eta'''\beta +
\bar\eta \bar\eta''' \bar\beta \biggr \},
\end{eqnarray}
%
where
\begin{eqnarray}
\beta_{\mu\nu}(x) = \sum\limits_{n=0}^\infty \hbar^n
\;\beta_{\mu\nu}^{(n)}(x),
\nonumber
\\
\beta = \sum\limits_{n=0}^\infty \hbar^n \beta^{(n)}, \quad
\bar\beta = \sum\limits_{n=0}^\infty \hbar^n \bar\beta^{(n)}.
\end{eqnarray}
%
Here $\beta^{(n)}_{\mu\nu}(x)$ are arbitrary functions and $\beta^{(n)},
 \bar \beta^{(n)}$ are the arbitrary constants. Thus we see that the
general form of anomaly (41) can be established up to above functions
and constants using only the generic capabilities of generalized canonical
quantization. Certainly, to find these functions and constants we should
perform some kind of loop calculation. The example of such calculations
was given in ref.[18]. It is evident that the found general form of
anomaly (41) in any case allows to simplify the process of real loop
calculations.

      6. Now we are going to investigate an arbitrariness in the
anomaly structure (41). Let ${\cal A}$ be a solution of the equation
$\delta {\cal A} = 0$. Then $\tilde {\cal A} = {\cal A} + \delta \Theta$
with any $\Theta$ is also a solution of this equation. Writing
\begin{eqnarray}
\tilde {\cal A} = i\hbar^2\;\int d\sigma \tilde a, \quad
{\cal A} = i\hbar^2\;\int d\sigma a, \quad
\Theta = i\hbar^2\;\int d\sigma \theta,
\nonumber
\end{eqnarray}
we obtain
\begin{eqnarray}
 \tilde a =  a + \delta \theta + \partial_\sigma \rho
\nonumber
\end{eqnarray}
%
with some $\rho$. Here $\dim \theta = 2 + c, \dim \rho = 2 + 2c$. Hence
\begin{eqnarray}
(\eta + \bar\eta)(\eta - \bar\eta)'\;\tilde{\beta}_{\mu\nu}(x) Y^\mu
\bar Y^\nu + \eta \eta'''\tilde{\beta} + \bar\eta \bar\eta'''
\tilde{\bar{\beta}}  = {}
\nonumber
\\
{} = (\eta + \bar\eta)(\eta - \bar\eta)'\;\beta_{\mu\nu}(x) Y^\mu \bar Y^\nu +
\eta \eta'''\beta  +
\bar\eta  \bar\eta'''  \bar\beta  +  \delta \theta + \partial_\sigma \rho.
\end{eqnarray}
%

      The dimensions of $\theta$ and $\rho$ allow to write
\begin{eqnarray}
\theta&=&\eta M + \bar \eta \bar M + \eta' N + \bar \eta' N +
\eta'' P + \bar \eta'' \bar P \quad\,
\nonumber
\\
\rho&=&\eta\eta'K + \bar \eta \bar \eta'\bar K + \eta \bar \eta L +
\eta' \bar \eta Q + \bar \eta' \eta' \bar Q +
\nonumber
\\ & &
+ \eta\eta''V +
\bar\eta\bar\eta''\bar V + \eta''\bar\eta Z + \bar\eta''\eta'\bar Z,
\end{eqnarray}
%
where $\Phi \equiv (M, \bar M, N, \bar N, P, \bar P)$ and
$\chi \equiv (K, \bar K, L, Q, \bar Q, V, \bar V, Z, \bar Z)$ are some
functions with definite dimensions.

       A straightforward
calculation shows that $\partial_\sigma \rho$ have the ghost
structure given by the right hand side of the eq.(43) only if $\chi = 0$ and
hence $\rho = 0$.
Now let as consider the structure of $\delta \theta$. According to the
eq.(44) one gets $\dim M = \dim \bar M = 2, \dim N = \dim \bar N = 1$
and $\dim P = \dim \bar P = 0$. Therefore $M, \bar M$ can be expanded
in the basis of $ Y^\mu \bar Y^\nu, Y^\mu Y^\nu, \bar Y^\mu \bar Y^\nu, Y'^\mu,
\bar Y'^\nu $; the $N, \bar N$ can be expanded in the basis of $Y^\mu, \bar
Y^\mu$ and $P, \bar P$ are arbitrary functions of $x$. Taking into
account the properties of $\Phi$, calculating $\delta \theta$ and
demanding that $\delta \theta$ has the ghost structure given by the right hand
side of eq.(43) one gets
\begin{eqnarray}
\theta &=&
\eta \left( -{1\over 18}\partial_\mu \xi_\nu Y^\mu \bar Y^\nu +
{1 \over 18}\partial_\mu \xi_\nu Y^\mu Y^\nu -
{1\over  9}\xi_\mu Y'^\mu \right) + {}
\nonumber
\\
&+& \bar\eta \left( -{1\over 18}\partial_\mu \xi_\nu Y^\mu \bar Y^\nu + {}
{1\over 18}\partial_\mu \xi_\nu \bar Y^\mu \bar Y^\nu +
{1\over 9} \bar \xi_\mu \bar Y'^\mu \right) - {}
\nonumber
\\
&-& {1\over 3} \eta' \xi_\mu Y^\mu + {1\over 3} \bar \eta' \xi_\mu \bar Y^\mu ,
\end{eqnarray}
%
where $\xi_\mu(x)$ is an arbitrary vector
field satisfying the relation $\nabla_\mu \xi_\nu (x) = \nabla_\nu
\xi_\mu (x)$. Here $\nabla_\mu \xi_\nu (x)$ is a covariant derivative
in terms of background metric $G_{\mu\nu} (x)$.

      The eqs.(43, 45) together with $\rho = 0$ lead to
\begin{eqnarray}
\tilde{\beta}_{\mu\nu}(x)&=&\beta_{\mu\nu}(x) +  \nabla_\mu \xi_\nu (x),
\nonumber
\\
\tilde \beta&=&\beta, \quad
\tilde{\bar{\beta}} = \bar\beta.
\end{eqnarray}
%

The eqs.(46) define the admissible ambiguity in general structure of the
anomaly (41).

       7. We have considered the problem of anomaly in string theory
interacting with background gravitational field in framework of
generalized canonical quantization procedure. This problem
has been formulated in terms of symbols of operators as a problem of
solution of the equation $\delta {\cal A}(\eta, \bar \eta, x, p) = 0$
where $\delta$ is the canonical BRST - transformation (5) and ${\cal A}$
is a symbol of the anomaly operator. A general solution of above equation
has been obtained and an admissible arbitrariness in the solution has
been described. We see, in principle, that the anomaly problem is
reduced to a cohomology problem for the operator $\delta$ (5) acting in
space of c-valued functions depending on coordinates of the extended
phase space.

       Taking into account the general structure of the anomaly (41) one
can immediately write the conditions of anomaly cancellation in the
form $\beta_{\mu\nu} (x) = 0, \beta = 0, \bar \beta = 0$. As we have already
mentioned the functions $\beta_{\mu\nu}(x)$ and the constants $\beta, \bar
\beta$
are not defined within the approach under consideration. Therefore,
 we should develop a calculational technique allowing to compute
the function $\beta_{\mu\nu}$ and the constants $\beta, \bar \beta$
only on the base of the generalized canonical quantization scheme. The
first step towards such a technique was undertaken in ref.[18].

      The approach general and allows to
investigate various string models in background fields. In
particular, inclusion of massless antisymmetric tensor field in
to the theory will not lead to any problems and will not demand
essential modification of the formalism.

     {\bf Acknowledgments.}

The  research described in this paper has been supported in parts by
International Science Foundation, grant N RI1000 and Russian Basic Research
Foundation, project N 94-02-03234.

\vspace{0.5cm}
   {\bf References. }

\begin{enumerate}
\item
Lovelace C. Phys. Lett. 135B (1984) 75.
\item
Fradkin E.S., Tseytlin A.A.  Phys. Lett. 158B (1985) 316;
Nucl. Phys. B261 (1985) 1.
\item
Callan C., Martinec E., Perry M., Friedan D.  Nucl. Phys. B 262 (1985) 593.
\item
Sen A. Phys. Rev. D32 (1985) 210.
\item
Tseytlin A.A. Nucl. Phys. B294 (1987) 583.
\item
Osborn H.  Nucl. Phys. B294 (1987) 595.
\item
Tseytlin A.A. Int. J. Mod. Phys. A 4 (1989) 4249.
\item
Osborn H. Ann. Phys. 200 (1990) 1.
\item
Fradkin E.S., Vilkovisky G.A. Phys. Lett. 55B (1975) 224.
\item
Batalin I.A., Fradkin E.S. Phys. Lett. 128B (1983) 303;
Nucl. Phys. B279 (1989) 514.
\item
Batalin I.A., Fradkin E.S. Riv. Nuovo Cim. 9 (1986) 1.
\item
Batalin I.A., Fradkin E.S. Ann. Inst. Henri Poincare 49 (1988) 145.
\item
Kato M., Ogawa K. Nucl. Phys. B212 (1983) 443.
\item
Hwang S. Phys. Rev. D28 (1983) 2614.
\item
Marnelius R. Nucl. Phys. B211 (1983) 14.
\item
Marnelius R. Nucl. Phys. B221 (1983) 409.
\item
Henneaux M., Teitelboim C. Quantization of Gauge System. Princeton
University Press; Princeton, New Jersey, 1992.
\item
Buchbinder I.L., Fradkin E.S., Lyakhovich S.L., Pershin V.D.
Int. J. Mod. Phys. A6 (1991) 1211.
\item
Fujiwara T., Igarashi Y., Kubo J., Maeda K. Nucl. Phys. B391 (1993)
211.
\item
Fujiwara T., Igarashi Y., Koseki M., Kuriki R., Tabei
Nucl. Phys. B425 (1994) 289.
\item
Berezin F.A., Shubin M.A. Schr\"{o}dinger Equation. Moscow
University Press; Moscow, 1983 (in Russian)
\end{enumerate}

\end{document}